# Testing the 10 spectrograph units for DESI: approach and results

S. Perruchot[a], P.-E. Blanc[a], J. Guy[b,f], L. Le Guillou[b], S. Ronayette[a], X. Régal[a], G. Castagnoli[a], A. Le Van Suu[a], E. Sepulveda[b], E. Jullo[c], J.-G. Cuby[c], S. Karkar[b], P. Ghislain[b], P. Repain[b], P.-H. Carton[g], C. Magneville[g], A. Ealet[d], S. Escoffier[d], A. Secroun[d], K. Honscheid[e], A. Elliot[e], P. Jelinsky[f], D. Brooks[i], P. Doel[i], Y. Duan[j], J. Edelstein[f], J. C. Estrada[k], E. Gastañaga[l], A. Karcher[f], M. Landriau[f], M. Levi[f], P. Martini[e], P., N. Palanque-Delabrouille[h], F. Prada[m], G. Tarle[n], K. Zhang[f], for the DESI collaboration

[a]Aix-Marseille Université, CNRS, Institut Pytheas-Observatoire de Haute Provence, 04870 St-Michel-l'Observatoire, France; [b]Sorbonne Universités, UPMC Université Paris 06, Université Paris-Diderot, CNRS/IN2P3 LPNHE 4 Place Jussieu, F-75252, Paris Cedex 05, France; [c]Aix Marseille Univ, CNRS, CNES, LAM, 13388 Marseille, France; [d]Aix Marseille Univ, CNRS/IN2P3, CPPM, Marseille, France; [e]Department of Physics, The Ohio State University, 191 West Woodruff Avenue, Columbus, OH 43210, USA; [f]Space Sciences Laboratory, University of California, Berkeley, 7 Gauss Way, Berkeley, CA 94720, USA; [h]CEA, centre deSaclay, RFU/DPhP, F-91191 Gif-sur-Yvette, [i]University College London, UK, [j]Boston University, Boston, MA, [k]Fermilab, University of Chicago, IL, [l]Institute of Space Sciences (ICE, CSIC), Barcelona, Spain, [m]Instituto de Astrofisica de Andalucia CSIC, Granada, Spain, [n]University of Michigan, Ann Arbor, Michigan

## ABSTRACT

The recently commissioned Dark Energy Spectroscopic Instrument (DESI) will measure the expansion history of the Universe using the Baryon Acoustic Oscillation technique. The spectra of 35 million galaxies and quasars over 14000 sqdeg will be measured during the life of the experiment. A new prime focus corrector for the KPNO Mayall telescope delivers light to 5000 fiber optic positioners. The fibers in turn feed ten broad-band spectrographs.

A consortium of Aix-Marseille University (AMU) and CNRS laboratories (LAM, OHP and CPPM) together with LPNHE (CNRS, IN2P3, Sorbonne Université and Université de Paris) and the WINLIGHT Systems company based in Pertuis (France), were in charge of integrating and validating the performance requirements of the ten full spectrographs, equipped with their cryostats, shutters and other mechanisms.

We present a summary of our activity which allowed an efficient validation of the systems in a short-time schedule. We detail the main results. We emphasize the benefits of our approach and also its limitations.

## 1. INTRODUCTION

The Dark Energy Spectroscopic Instrument [1] (DESI) is a Stage IV ground-based dark energy experiment that will produce an unprecedented three-dimensional map of the Universe. DESI employs the baryon acoustic oscillation (BAO) technique to provide state-of-the-art cosmological constraints to redshifts > 2 by observing several tracer galaxy populations [2]. In total, approximately 35 million redshifts will be obtained by DESI during its five-year survey which will cover nearly the entire northern (δ >−30°) sky at high Galactic latitudes [3] (∼14,000 square degrees in total).

Importantly, DESI will collect redshifts for ~15 million faint ($g$~23 AB) emission line galaxies at $z$~1−1.5, a redshift range largely unexplored with BAO.

DESI consists of a next-generation multi-object spectroscopy instrument ([4], [5]) installed at Kitt Peak National Observatory's Nicholas U. Mayall 4-meter Telescope in southern Arizona. DESI's ten spectrographs combine to acquire 5,000 spectra simultaneously, spanning the 360-980 nm wavelength range with resolution $\lambda/\Delta\lambda \sim 2,000-5,500$. DESI's focal plane resides at the prime focus of the Mayall telescope, where a new DESI top end [6] has been installed. The DESI corrector [7] provides a large 3.2° diameter field of view, of which ~7.5 square degrees is instrumented for spectroscopy. A hexapod allows for fine-grained adjustments of the corrector barrel position. 5,000 fiber positioning robots ([8], [9]) patrol the focal plane to align stars and galaxies with fiber-optic cables connected to the spectrographs. DESI installation [10] completed in October 2019, and the Mayall facility will be entirely dedicated to DESI operations until the survey's completion.

This paper is dedicated to the spectrographs' validation. After a brief description of the spectrograph in section 2, the tests, resources and tools are presented in section 3. The main results for all of the ten spectrograph units are summarized in section 4 and we discuss the benefits and limits of our approach in section 5.

## 2. SPECTROGRAPHS BRIEF DESCRIPTION

### 2.1 Spectrograph brief description

The DESI spectrographs are fully described in [11]. The optical layout and global implementation is shown in Figure 1. A DESI spectrograph receives the light from 500 fibers. Ten spectrographs are used to accommodate the 5000 fibers of the DESI instrument. They are installed inside a temperature-controlled enclosure in the Coudé room at the Mayall telescope, with a variation of temperature of less than ±0.5°.

In each spectrograph, the 500 fibers are rearranged in the shape of a curved slit. The light is collimated by a spherical mirror, then separated by means of two dichroic windows into three channels (blue, red, near infrared) covering the 360 - 980 nm spectral range. In each channel, volume phase holographic gratings provide light dispersion. The resulting spectrum is then focused using a 5-lens camera, on a 4K x 4K, 15 μm pixels CCD detector, actively cooled down to 140 K (Red and NIR) or 163 K (Blue). A mechanical shutter in front of the slit is used to control the exposure time [12]. LEDs mounted onto the shutter and directed to the slit can be switched on to back illuminate the fibers. This feature is used for measuring the positions of the fibers in the telescope's focal plane. In order to avoid strong signal on the near-infrared CCD, this camera has its own shutter, which remains closed when this fiber position measurement is being performed. The spectrograph is also equipped with a pair of remotely controlled Hartmann doors situated in front of the collimator mirror for focus adjustment.

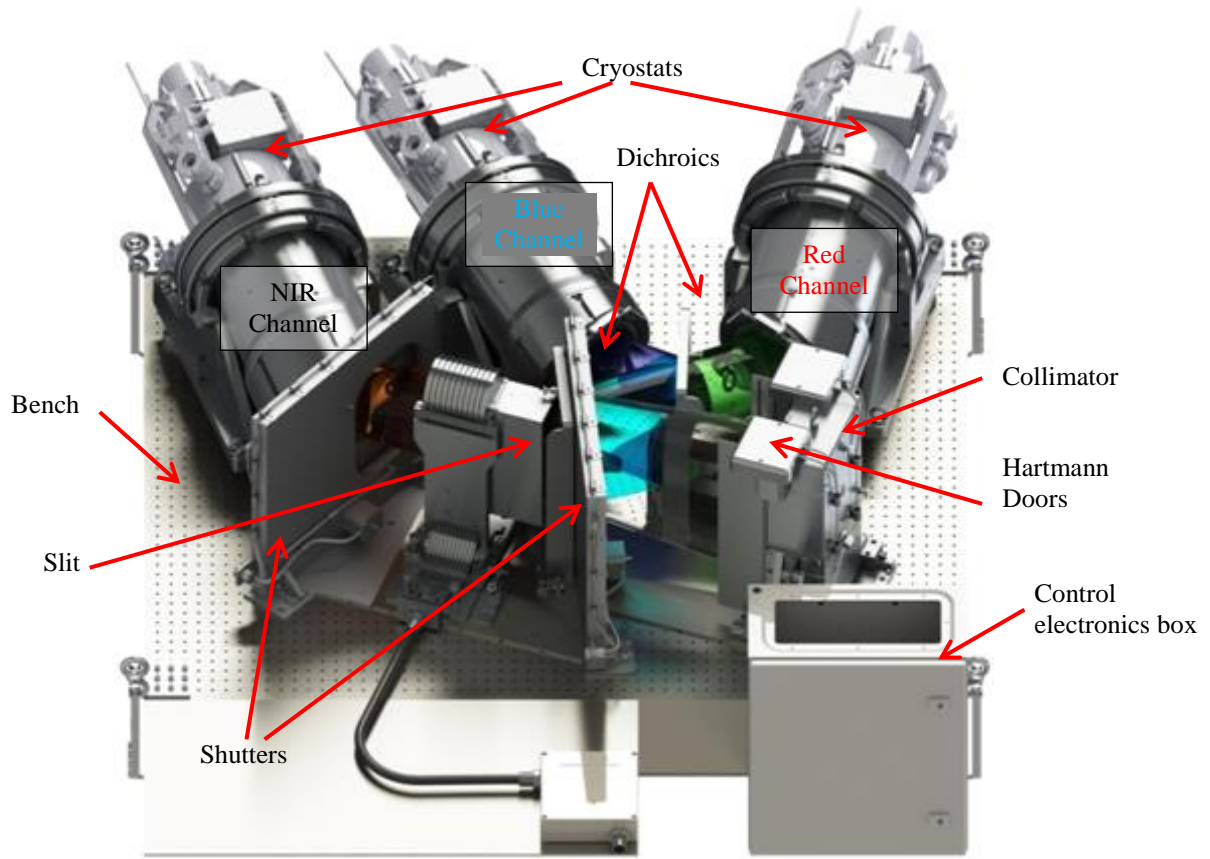

Figure 1. Mechanical implementation of one DESI spectrograph (1.8 m wide × 1.4 m deep × 0.6 m high).

## 2.2 Spectrograph design performance summary

The magnification of each camera is about 0.5, resulting in a fiber spot size onto the detector of about 50 μm, corresponding to 3 pixels. The cameras have an rms spot radius of less than 12 μm, providing allowance for manufacturing and alignment tolerances while keeping the spot size limited by the projected size of the fiber core (107 μm diameter). Table 1 below summarizes the performance requirements of the spectrograph.

Table 1. Main spectrograph performance requirements.

| Channel | Spectral range | Resolution | End-to-end throughput |
|---|---|---|---|
| Blue | 360-593 | 2000 to 3200 | $\lambda$=400 nm: 60%<br>$\lambda$=500 nm: 69% |
| Red | 566-772 | 3200 to 4100 | $\lambda$=600 nm: 69%<br>$\lambda$=700 nm: 70% |
| NIR | 747-980 | 4100 to 5100 | $\lambda$=800 nm: 78%<br>$\lambda$=900 nm: 73% |

# 3. SPECTROGRAPH TESTS

## 3.1 Spectrograph test plan rationale

The spectrograph tests come as integration and validation test after a substantial amount of testing of individual components by the providers. For example, Winlight tested the optical part of the spectrograph, CEA the cryostats, LBL tested the gratings, dichroics, and FEEs, and OSU tested the mechanisms and control electronics.

The spectrograph test plan rationale has been described in [13] and [14]. First, an engineering model (EM) was built and fully tested before starting the production of the remaining nine spectrographs and minor retrofitting modifications to transform the EM unit to the first spectrograph SM1. Then, the test plan for the spectrograph qualification was simplified twice; a first time for SM1, SM2 and SM3, and a second time for all the others, to speed up the delivery schedule, but trying to mitigate risks. See Table 2 to find the list of the tests performed during all these test phases.

## 3.2 Test tools

The tools used all along in the test phases have already been described in [14]:

- A sparse test slit, in place of the 500 fibers slit that feed the spectrograph at the telescope. This sparse slit consists of 21 fibers rearranged in a slit mechanically identical to the final DESI slit. Note that another sparse slit is used at Mayall telescope. See [15] for details.

- A fiber illumination tool, specifically developed by AMU [14], to feed the sparse test fiber slit, either some fibers individually or all at once, with the following light sources: four spectral lamps (Kr, HgAr, Cd, Ne), a tungsten lamp (continuous spectrum) and 6 LEDs that provide 20 nm wide powerful spectra.

- A throughput tool, especially developed by LPNHE [14], to perform an absolute measurement of the flux at the exit of each fiber of the sparse test slit fed by the fiber illumination tool (entrance of the spectrograph) and compare it to the absolute flux measured by the spectrograph, with the help of a calibrated photodiode.

- A flat field slit, which illuminates the entire instrument pupil with a continuous slit - see [15]

The Fiber Illumination Tool system is operated with the DESI Instrument Control System (ICS), enabling automation of the measurements sequences and remote operation, and allows the current parameters used to be in the FITS headers. The ICS was also tested and improved during this period.

Table 2. Test list as done.

| Test designation | EM phase | SM1 | SM2-SM3 | SM4 ... – SM10 | SM1-SM10 Mayall |
|---|---|---|---|---|---|
| Focus | | | | | |
| Full-Pupil Focus | + | + | + | + | + |
| Hartmann Doors focus scan | + | + | + | + | |
| Optical performance | | | | | |
| Image quality | + | + | + | + | * |
| Trace coordinates and wavelength solution, resolution | + | + | + | + | + |
| Imaging and spectral ranges | + | + | + | + | |
| Flat field | + | + | SM3 | | |
| Spectrograph throughput | + | + | + | + | |
| CCD properties | | | | | |
| Readnoise | + | + | + | + | |
| CCD gains | + | + | + | | |
| Amplifier cross-talk | + | + | + | + | |
| Straylight | | | | | |
| Narcissus effect | + | partial | + | | |
| Light leaks | + | | + | | |
| 2nd order contamination | + | | + | | |
| Fiber to fiber contamination | + | + | + | + | |
| Functional tests | | | | | |
| Shutters | + | + | + | + | |
| Hartmann doors | + | + | + | + | |
| Fiber illuminator | + | | | | |
| NIR shutter light tightness | + | | + | | |

### 3.3 Analysis tools

All the analyses were done offline, with a combination of custom Python scripts and prototype DESI pipeline routines. Some tools are based on BOSS experience (Preprocessing, Wavelength solution), and others were specifically developed. The specific throughput analysis is fully described in [14]. The analyses tools needed some adaptations and optimizations during the tests phases, but we tried to maintain a homogeneous analysis along the tests of all units. We describe here the two main tools.

Focus analysis tool

The first steps, focusing and quick image quality analysis (good health), have to be done without knowing much about the system. We wrote a custom Python script, which was extensively used to analyze the dedicated images, made in each channel with a different set of spectral lamps (see section 4.2). This script consists in 5 steps:

- simple preprocessing
- selection of 80-100 spots in each channel, imaging well-distributed spectral lines across each field of view (all fibers fed), on a grid (to ease data interpretation with wavelengths and fibers identification) - see Figure 8
- extraction of spot characteristics: centroid position and FWHMs along the X and Y directions for each focus scan position. Each PSF (Point Spread Function) is fitted with a 2D Gaussian. We use a normalization method to account for the fact that the spectral lines have different fluxes. We choose the Noise Equivalent Area (NEA), an effective PSF area in pixel units: two PSFs with the same NEA will give the same signal to noise ratio for an emission line if the CCD pixel noise is dominated by readout noise
- best defocus determination for each spot, as the minimum of the fitting parabola
- best plane derivation. We fit a tilted plane to the best focus as a function of position on the detector and use the results to adjust the tip/tilt and focus position of each cryostat.

As the FWHMs are calculated for all spots in any image, this code also gives access to the image quality with one image taken at the best plane.

Wavelength solution

The spectral resolution is assessed in two steps. We first take continuum lamp spectra to determine the trace of the spectra. Then we take spectral lamp images to determine the wavelength solution. We are then able to derive the spectral resolution.

The initial wavelength solution is found by matching triplets of spectral lines measured (in pixel units) with known triplets of lines (in wavelength units) from the lamp's lines list. Once a first solution is found, the full solution is determined with more lines and higher order polynomial coefficients. Subsequent adjustments are determined by cross-correlation with this full solution. To reduce the effect of noise and to reject cosmic rays we combine several images of identical settings and we measure the PSF shape modeled from a basis of polynomials convolved with Gaussians. We make use of the NEA normalization method to account for the fact that the spectral lines have different fluxes along the dispersion direction.

### 3.4 Schedule and human resources

The whole spectrograph qualification phase lasted from the end of 2016 until January 2020. The Engineering model (EM) was tested from October 2016 to July 2017, the production series of 10 spectrographs (including modified EM) were tested in France from March 2018 to October 2019 (20 to 30 days for performance tests), and the spectrographs refocusing at their final location at the Mayall Observatory was performed from February 2019 to January 2020.

Each spectrograph model was tested with all its components, including the cryostats (including CCDs, Front End Electronics or FEEs), shutters, Hartmann Doors, fiber illuminators and electronics cabinet. An OSU team came to set up the mechanisms for the EM unit and trained the local team to do this for the remaining spectrographs. The three cryostats were attached to the spectrograph and tested for good health behavior by a CEA team (3 persons) coming especially to the test place for this delicate operation. Then a system configuration was requested and provided remotely by the ICS team, with the help of the local AMU team. After some functional tests, especially readnoise behavior, the performance tests were performed. The preliminary phase usually lasted about 2 weeks. It can be noted that this required tight interaction between several distant teams, and depended on some key persons availability.

The performance tests begin with an interactive phase: the focusing, the best plane being requested for all the other performance tests. The robust but long procedure we choose is described in details in section 4.2. The focusing phase for the three cameras of each spectrograph usually mobilized two or three persons near full time during about two weeks.

Some of the analyses were derived from the wavelength solution, such as the spectral and imaging ranges. Others required the development of specific software tools prepared by the Data Reduction Software (DRS) team, such as readnoise analysis, or a spectral extraction to characterize the fiber-to-fiber contamination for instance. This early data set has proven to be very useful for the development of the DESI spectroscopic pipeline.

All along the project, a close interaction between the technical teams and the scientists was necessary and beneficial to all.

### 3.5 Images

A large amount of images was recorded, from 1,000 to 3,000 depending on the spectrograph during the performance series phase, and about 23,000 in total. Not all these data have been analyzed, among them there is a lot of dark images. Most images have been acquired for a specific purpose, but could be used later on for another. For instance this has proven useful to understand an issue found with the collimator coating (see section 5.2)

For SM4 to SM10, all the analyses have been re-optimized for a faster validation of the spectrographs.

### 3.6 Encountered (and mostly solved) issues

Unsurprisingly, some problems occurred during the integration of these ten units, each consisting of multiple subsystems. All problems were solved without degrading the performance of the spectrographs nor blocking the qualification tests. However, they induced delays. As a lesson learned, we would recommend to put even more efforts to the validation of the sub-systems prior to their shipment for integration. More documentation and installation procedures would also have eased the integration and testing.

Optical issues.

Two VPH gratings were mounted upside down on EM, due to a misinterpretation of the marks on the optics. This was thankfully detected with the throughput test and easily corrected for all the series units.

On the second spectrograph, the collimator had to be realigned because one channel was not satisfying the required optical quality. But because the teams were co-located, this caused only a small delay by switching to another spectrograph.

An error in the mounting of a cryostat window lens on the cryostat lead to a cryostat swap, which also implied additional delays. This occurred because the spectrographs were not aligned with the final cryostat window lens by the manufacturer. This error also happened because the lens coating made the visual check very difficult. A more reliable test could have been introduced.

Fluctuations in the positions of spots on the CCD were detected during the analysis of the focus scan data. The issue was traced back to backlash in the piston screws. The solution was to pay more attention during the mechanical operations, to perform a new scan if necessary, and to always make the scans in only one direction to avoid such jumps in the spot position on the movement back to central position. The three position gauges (less than 1 micron accuracy) were essential to achieve the whole focus process, including tilt correction.

The collimators coating shows spectral absorption features in the blue region around 440 nm. The features are pronounced for all spectrographs but the first one (SM1). It was not detected before the installation at Mayall Observatory. This defect could have been detected in Tungsten spectra but was not, because we were not aware of such a risk and the datasheets were globally compliant. The spectrographs still meet the science requirements. But we are working on replacing the nine collimators with new ones.

Cryostats issues.

After the tests with the EM spectrograph, a new system was put in place for SM1 and the rest of the series. Only twice, a small leakage was detected. The solution was to constantly dynamically pump (secondary pump), without challenging the test results quality.

Some problems required a cryostat swap and cryostat return to Saclay, and then implied delays: a temperature probe failure, an interlock wiring error. We would have benefited from more complete tests by the expert team before shipping or upon delivery onsite.

Cabling issues.

Some cabling issues happened, on Hartmann Doors and shutter cables, despite the initial testing by the provider, which may have been deeper. We had spares and this helped a lot, since we were able to swap. But time was lost before the detection of problems. A more efficient keying policy on the cables would have helped, since some mistakes happened. And a test bench on harnesses would have been useful.

<u>Software issues.</u>

Various software issues were solved remotely, with delays never exceeding 3 days: devices addresses mismatches, ICS update incompatibility, FEE and CCD configuration file mismatches. The different software systems evolved at different rhythms and were not automatically implemented on the performance test site. There was no reason to do so, but this sometimes added a debugging phase. Documentation, especially an installation manual, would have helped, since the implicated teams were distant.

Other minor problems during operations, requested only to restart programs and patience from operators.

## 4. SPECTROGRAPHS QUALIFICATION: MAIN RESULTS

### 4.1 All spectrographs validated

First, and the main result, is that all the ten spectrograph models have been validated, without major issue during the test phase, in less than 20 months for the series phase in France. Hereafter we present the very good consistency in the optical performances.

### 4.2 Focus

*Focus adjustment operation*

We manually adjust the focus and tilt of each channel by translating along three axes the corresponding cryostats that include the focal plane arrays and field lenses. The scan ranges vary: ±0.5 mm (step 0.05 mm) for the first scan around the nominal optical manufacturer best plane position, ±0.2 mm after tilt correction for verification if needed, and ±0.150 mm at Mayall observatory after shipping.

At each position, we take exposures of a few seconds with the HgAr and Cd lamps in the Blue channel, HgAr and Ne lamps in the Red channel and HgAr and Kr lamps in the NIR channel. At Winlight, 3 full pupil exposures and one per Hartmann door closed, only one full pupil at Mayall Observatory.

*Focus adjustment analysis*

We measure the Point Spread Functions (PSF) width in several locations across the field of view (FoV). The lamps provide well-distributed spectral lines across the different fields of view. The analysis procedure is described section 3.3.

The results are in good agreement with expectations, including the best focus variation as a function of X (slit) and Y (wavelength) coordinates (field curvature, residual chromatism) – see Figure 4 and Figure 5 as examples.

*Distance between the optical plane determined by the manufacturer and the best plane with cryostats*

The large range for the first scan is justified by the possible distance between the optical focal plane position determined by the manufacturer (with a test cryostat lens) and the actual position requested for the CCD (with the final cryostat lens), as can be observed on Figure 2. With our method, it is necessary to have sufficient parabola arms length to find the minimum. Furthermore, a change in the cryostat building between EM and further models added a systematic difference of about 0.1 mm. The two large requested piston values (+0.5 and -0.3) for the blue channel of SM7 and SM9, in opposite directions, are not problematic but stay unexplained.

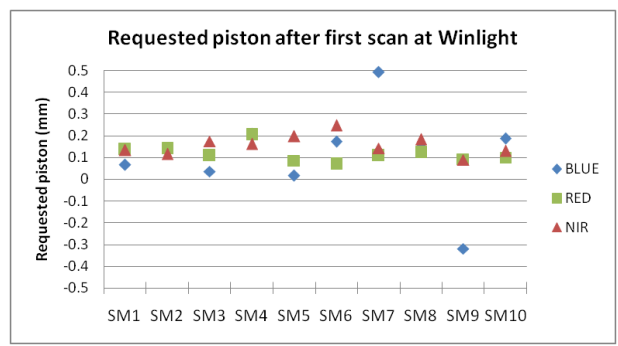

Figure 2. Mean piston determined by the first focus scan analysis, done on [-0.5mm;+0.5mm]. The reference position "0" is the focal plane position indicated by Winlight (without the cryostats).

*Piston between best plane at Winlight and best plane at Mayall*

The spectrograph models were shipped mounted, except for the cryostats that traveled separately. The focusing had to be redone considering also the change in altitude: about 200 m for the qualification tests in Pertuis at Winlight company (France) and 2120 m for Mayall Telescope. The test sparse slit was a new one, different from the one used in France. The pistons are very diverse as shown on Figure 3, from 10 μm extrafocal to 100μm intrafocal.

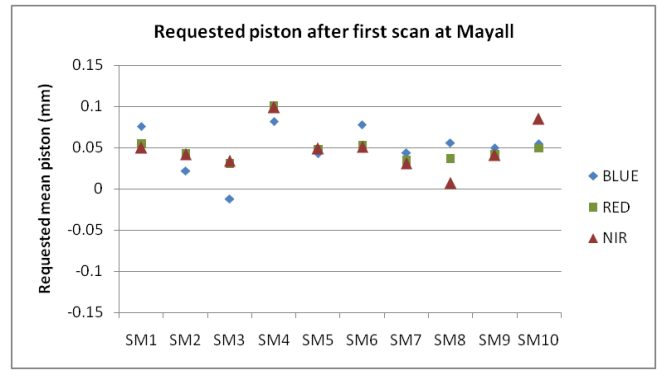

Figure 3. Mean piston determined after focus scan analysis at Mayall telescope, from qualification best plane in France. The reference position "0" is the focal plane position determined with the cryostats at Winlight.

*Tilt correction*

Figure 4 (for the Blue channel of SM9) and Figure 5 (for the Red channel of SM7) show the defocus determination as a function of fiber or wavelength for the first scan and after tilt correction. On those both cases, tilts are clearly visible, especially on imaging direction. The tilt correction is very efficient to put all points in the depth of focus. In case of high tilt, the field curvature may produce a miscalculation of the tilt to apply, but a new scan will always allow to find the right plane.

The amounts of tilt projected on spectral and imaging directions are shown for all spectrograph units on Figure 6. All were reachable to correction with the spectrograph mechanics. 0.1° corresponds to 0.1 mm piston on a side of the 61.44 mm-wide CCD. At Mayall Telescope, these amounts were considerably reduced and quite low but were nevertheless corrected (Figure 7).

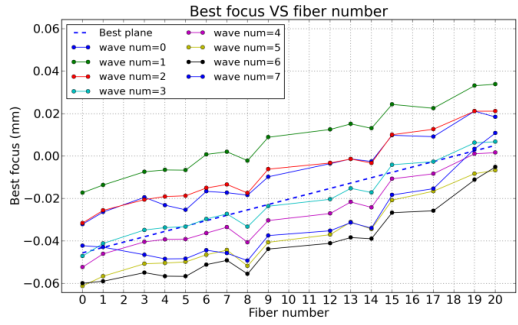
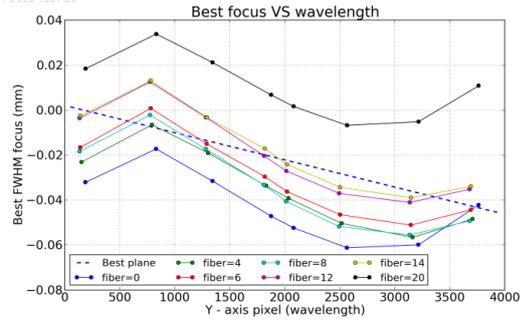

First scan

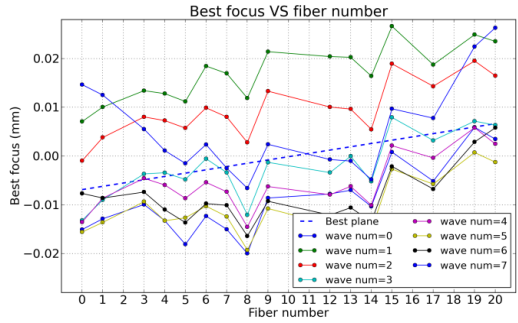
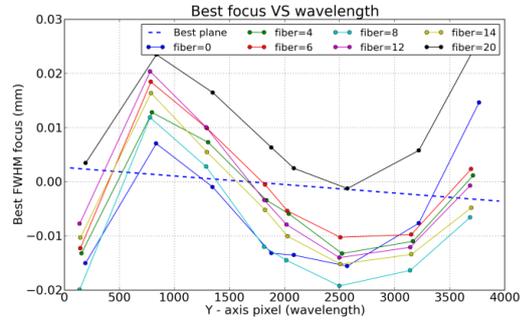

Scan after tilt correction

Figure 4. Blue channel on SM9: best plane defocus determination as a function of fiber (left) or wavelength (right), from first scan data (top) and after tilt correction scan data (bottom)

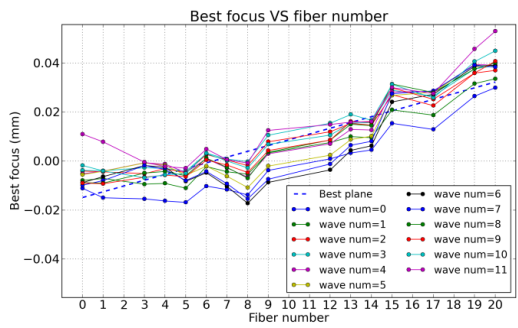
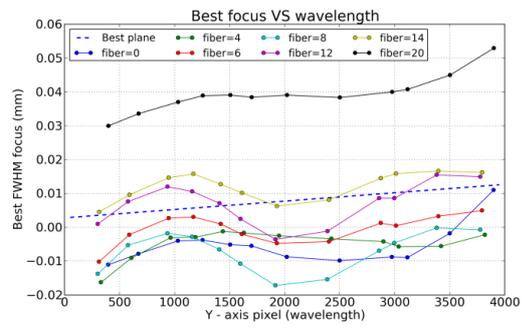

First scan

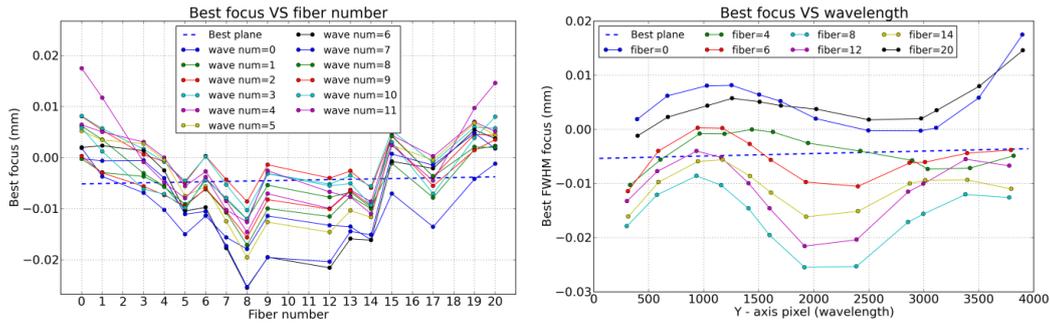
Scan after tilt correction

Figure 5. Red channel on SM7: best plane defocus determination as a function of fiber (left) or wavelength (right), from first scan data (top) and after tilt correction scan data (bottom)

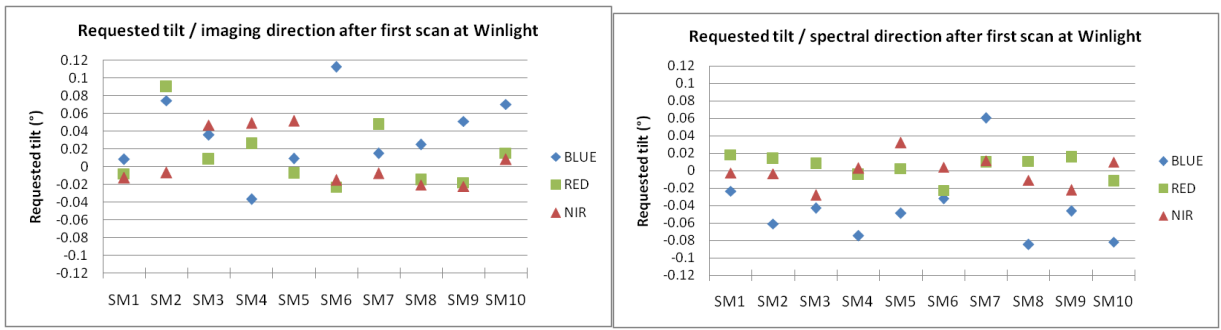

Figure 6. Requested tilts for all spectrographs after first focus scan during qualification tests.

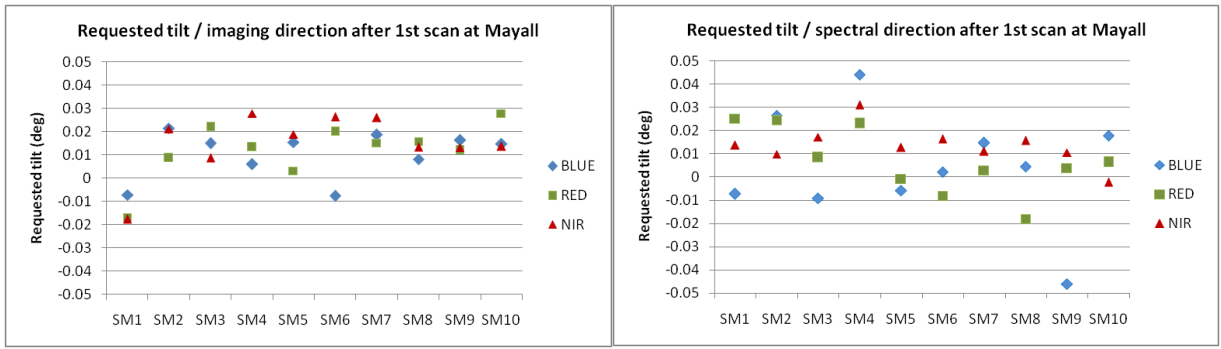

Figure 7. Requested tilts for all spectrographs after first focus scan at Mayall Telescope, from qualification best plane in France

### 4.3 Image quality

A quick image quality analysis is done before any sophisticated analysis and derivation, and uses short time exposures with spectral lamps when CCDs are at best plane: HgAr and Cd lamps in the Blue channel, HgAr and Ne lamps in the Red channel and HgAr and Kr lamps in the NIR channel.

The focus analysis tool is used to determine FWHMs along imaging and spectral dispersion directions. See an example of the results for SM7 Blue channel in Figure 9.

The image quality is estimated on the same points for all spectrographs (see Figure 8), with also points from known defocused fibers. The results are all compliant with the specifications and very similar from spectrograph to spectrograph except for the first of the series Blue SM1 (see Figure 10).

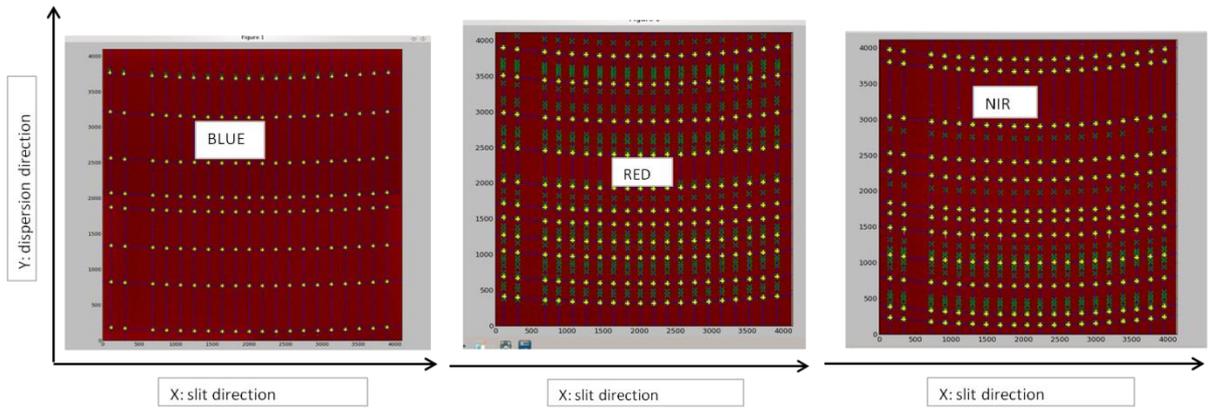

Figure 8. Selected sources for image quality estimation: Blue channel 160 points, Red channel 220 points, Nir channel 240 points.

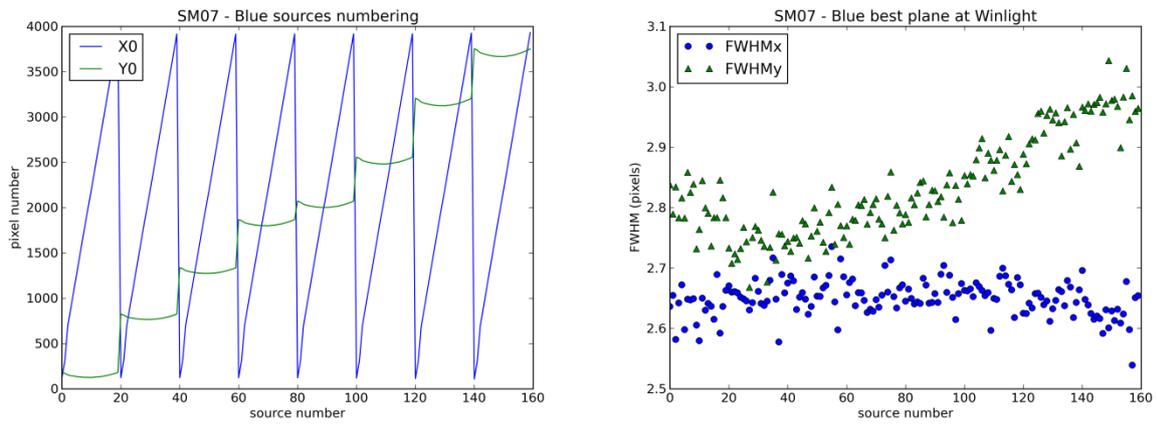

Figure 9. SM7's Blue channel best plane Image Quality quick analysis at Winlight. FWHMs derived for each source (right), the source numbering being explained by the coordinates X (imaging direction - slit) and Y (spectral direction) in pixels (left).

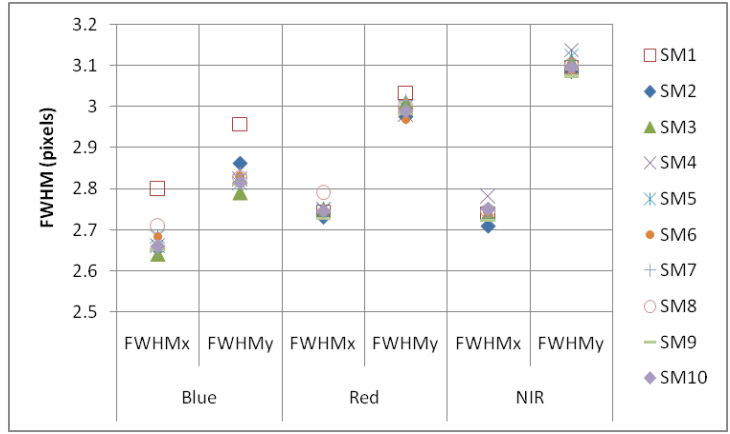

Figure 10. Mean FWHMs for all spectrographs, FWHMx along slit direction, FWHMy along spectral dispersion direction.

### 4.4 Wavelength solution, spectral resolution

Using the method described in section 3.3 (wavelength solution), we determine the spectral format for each fiber and find excellent agreement with the Zemax simulations, thereby validating the whole assembly, integration, alignment and verification procedure, including the positioning of the CCD inside the cryostat. Similarly, the wavelength dispersion (dλ/dpixel) as a function of wavelength and fiber in each channel is consistent with expectations from Zemax optical simulations.

The resolution measured from the PSF analysis (in the spectral direction), compared to the design (Zemax simulations) and the requirements, is shown on Figure 11 for SM9 as an example. We find that the resolution is slightly lower than expected from the design but fully meets the requirements over the full wavelength range. It is the case for all the spectrographs.

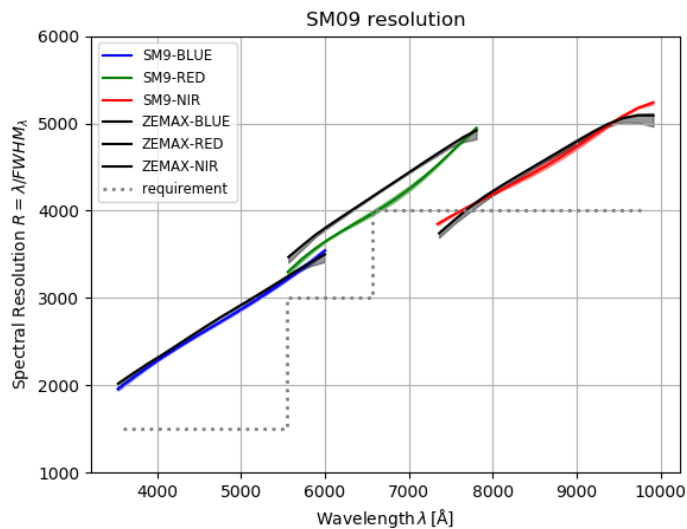

Figure 11 : SM9 resolution (in blue, green and ref for b, r and z channels) based on arc lamp exposures. To be compared with model (black).

### 4.5 Spectrograph throughput

The spectrograph throughput is the ratio of the amount of electrons collected on the CCDs for a known spectral flux at its entrance (fiber aperture). The throughput measurement procedure and analysis are fully described in [14].

The throughput measurement is done by using successively the 6 bright LEDs (370 nm to 940 nm) of the Fiber Illumination Tool, and by sequentially illuminating each single fiber of the sparse slit outside and inside the spectrograph. For a given LED/fiber configuration, we measure the fiber exit flux with the throughput measurement device, and we subsequently measure the flux in the corresponding spectrum onto the CCD. In both steps we monitor the illumination level using a second photodiode fixed on the integrating sphere of the Fiber Illumination Tool.

Figure 12 shows the measured throughput for SM8 compared with the estimate from the DESI optical model. The losses due to the fibers Focal Ratio Degradation (FRD) are not corrected for and they can be estimated around 10% for this first manufactured slit. Figure 13 summarizes the measurement result for each LED on all spectrographs, FRD not corrected. In addition to manufacturing differences, dispersion of throughput values among spectrographs could indicate some uncertainties in the measurements. Preliminary measurements on sky seem consistent with expectations.

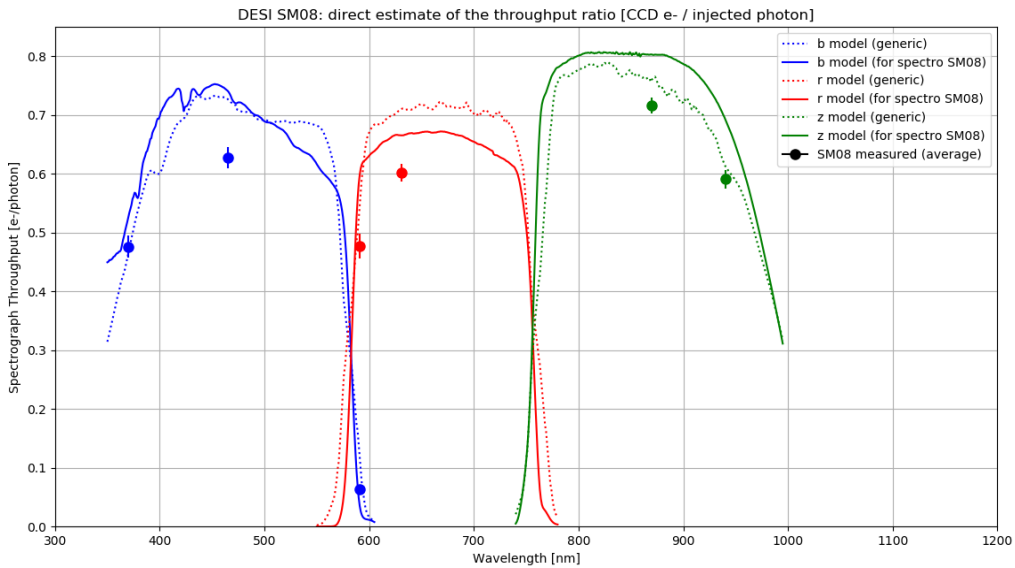

Figure 12 : Throughput measurements for SM8: measurements averaged on 13 test slit fibers. The error bars represent the fiber to fiber dispersion (RMS). On this graph, no FRD correction has been performed. The dotted line represents the throughput of the generic DESI optical model; the plain line represents the throughput calculation done using the throughput measurements from the separate optical elements for this specific spectrograph.

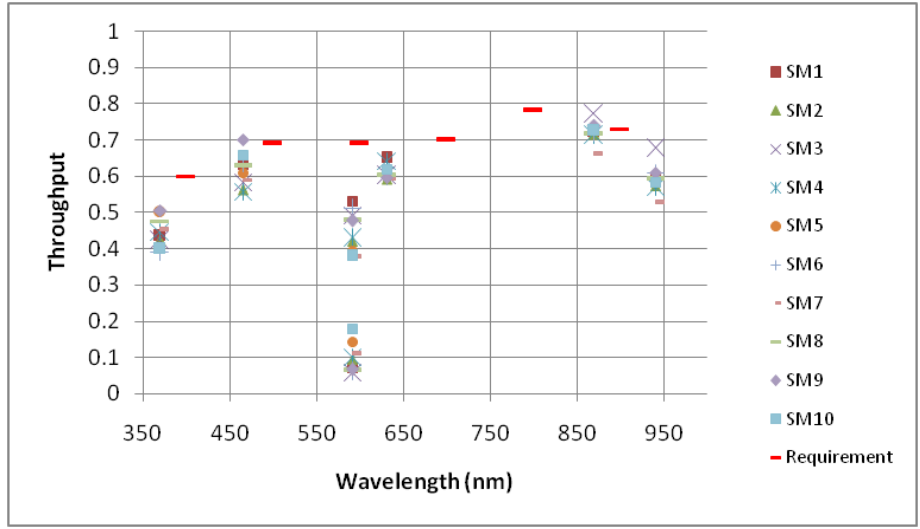

Figure 13. Measured throughput at LEDs wavelength for all spectrographs. No FRD correction.

### 4.6 Wavelength and imaging ranges, CCD positions

We use the trace of the spectra previously determined with the wavelength solution to derive the wavelength and spectral ranges. We verify that the whole slit is well imaged inside the CCD and that all the specified wavelength range per channel is inside the CCD. This is the case for all the spectrographs, as can be seen on Figure 14, where the minimal distance to the CCD side has been calculated. After SM1, the CEA responsible for the CCD's positioning in the cryostat (and other cryostat tasks) has improved its process. The reproducibility of the spectra corners positions is remarkable, regarding the spectrograph alignment and CCD positioning tolerances: on all channels and for spectrographs SM2 to SM10, a mean standard deviation in slit direction < 12 pixels (worst PTV 57 pixels) and in spectral direction < 7 pixels (worst PTV 32 pixels). Focal lengths of all spectrographs are very similar as dispersion powers are, as shown in Table 3, with differences less than 1% on image heights.

Table 3. Spectrum and slit image length on specified limits – spectrum length: distance between channel limit wavelengths for top or bottom fiber – slit length: distance between top fiber image and bottom fiber image at each channel limit wavelength.

|  | Blue channel [360-593nm] | | Red channel [566-772nm] | | Nir channel [747-980nm] | |
|---|---|---|---|---|---|---|
| **Length of the spectra** | **Bottom fiber** | **Top fiber** | **Bottom fiber** | **Top fiber** | **Bottom fiber** | **Top fiber** |
| Mean distance in pixels on all spectros | 3950.0 | 3949.4 | 3897.5 | 3900.5 | 3873.8 | 3875.1 |
| (max – min distance) /mean distance | 1.3% | 1.2% | 0.8% | 0.8% | 0.9% | 0.9% |
| **Length of the slit images** – distance btw top and bottom fibers at λ | **360 nm** | **593 nm** | **566 nm** | **772 nm** | **747 nm** | **980 nm** |
| Mean distance in pixels on all spectros | 3791.5 | 3828.6 | 3806.7 | 3827.6 | 3814.2 | 3827.4 |
| (max – min distance) /mean distance | 0.4% | 0.4% | 0.5% | 0.5% | 0.6% | 0.6% |

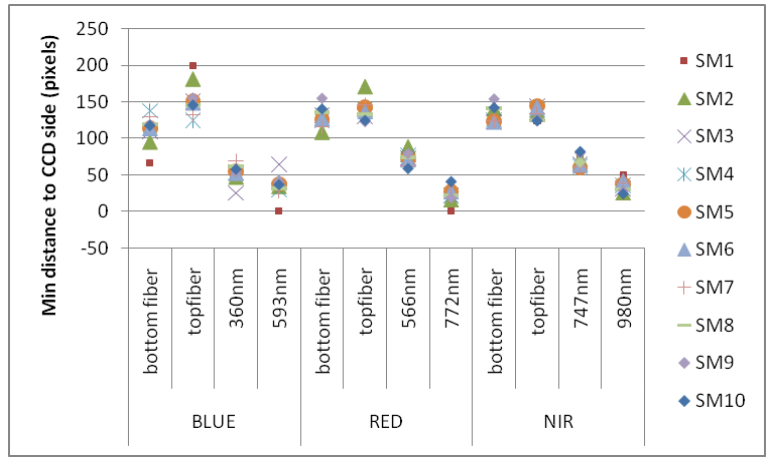

Figure 14. Minimal distance to CCD side for specified extreme wavelengths and extreme fiber positions in the slit for all spectrographs.

### 4.7 Straylight

Several types of straylight tests have been performed, and some of them very time consuming. Therefore, some of them have been descoped during production phase to speed up delivery, with a calculated risk (see Table 2) after the first models were tested.

Narcissus images.

Ghosts, or Narcissus images, are predictable images formed by undesirable back and forth reflections on the detector and optical surfaces. We track them by illuminating selected fibers at the center and edge positions along the slit with strong LED intensities and searching for faint secondary images on the images. As described in [14], we had an issue on EM1 due to an integration mistake, a ghost on the Blue channel. We only verified on some extra spectrographs (SM2, SM3, and partially on SM1) that the problem was well solved. Then the long test (long powerful LEDs exposures to reach a high dynamic between primary image and ghost) was no more done for further spectrographs.

Second order contamination

The rejection rate of the dichroics is not sufficient to block entirely the blue part of the gratings' second order in the Red and NIR cameras. To evaluate the level of second order contamination in these two channels, we illuminated the spectrograph with a strong blue LED signal and looked for the presence of second order spectra in the images with half an hour exposures (specifically, presence of second order spectra at 740 and 930 nm for blue LEDs emitting at 370 and 465 nm respectively).We did not find evidence of contamination by the second diffraction order in any of these settings, thereby validating the specifications and performance of the dichroics, for EM, SM2 and SM3 – see example Figure 15.

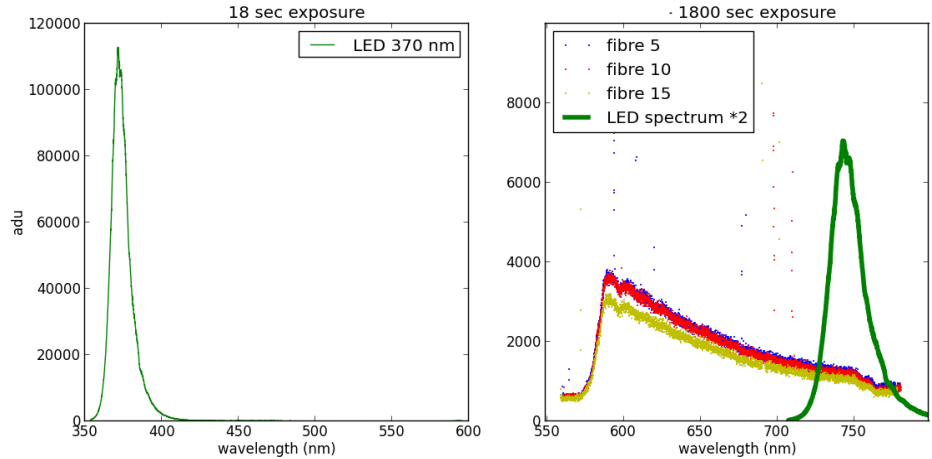

Figure 15: Second-order contamination test in SM2's RED channel. The 370nm LED's spectrum on Blue channel (left) could have produced a contaminating spectrum with doubled wavelengths on Red channel (calculated spectrum shown with arbitrary units in green on right figure). But no significant signal is seen where expected if significant second order contamination – see measured signals for 3 fibers in red, blue and yellow (right). The very low level slope observed is not due to second-order contamination.

Fiber-to-fiber cross-talk

The fiber-to-fiber crosstalk analysis has been done on all spectrographs, benefiting from the Tungsten images recorded for other purposes. It consists in looking at the signal at the location where an adjacent fiber would be (but is not, because we only use a sparsely populated test slit). Figure 16 presents the typical results obtained, with cross-talk always < 0.001 for Blue and Red Channels, and for Nir channel for wavelength < 915 nm typically (from 905 to 930 nm depending on the spectrograph unit). This level of fiber-to-fiber cross-talk is still acceptable on the entire domain and is anyway corrected for.

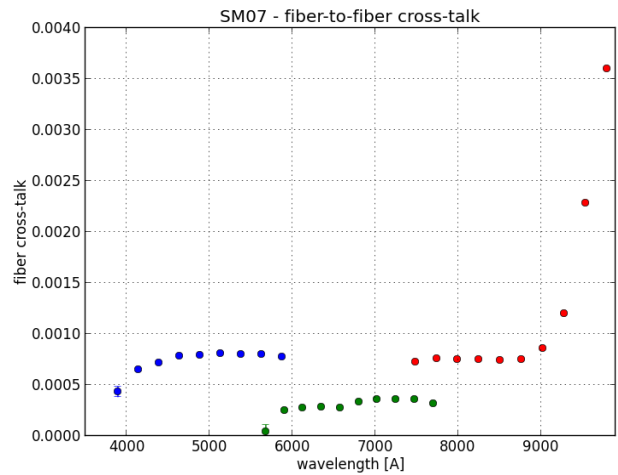

Figure 16: Fiber-to-fiber cross-talk on the three channels. Fraction of light from one fiber going into the next.

Light leaks.

A light leaks qualitative test has been performed for the EM, SM2 and SM3 spectrographs. The spectrograph was in nominal configuration, with all the covers' screws on and black silicon (RTV 133) on bellow/camera interfaces. Different configurations were set, with different type of ambient lights. The images from the detectors showed detectable levels of light, so that a precaution was taken later and for all tests to work in dark rooms (also in operation at the observatory).

**4.8 Detectors**

Detectors performance are dependent on FEE and CCD configurations. We tested some properties during these qualification phase, optimized some procedures or analysis code, but the real effort on detectors optimizing and characterizing has been done at Mayall Observatory.

Readnoise and amplifier crosstalk were verified for all the spectrographs, but not the gains or the flat field (only for EM, SM2 and SM3).

## 5. BENEFITS AND LIMITATIONS OF THE APPROACH

**5.1 Benefits**

Delivery schedule efficiency.

Our approach allowed the testing and validation of ten spectrographs with three channels each in about 20 months. It stood on an anticipated preparation phase for test tools, on an enriched prototype phase with a deep test plan, and on small stable teams with cumulative expertise, close interactions and reactivity.

Prototype phase

The quite long prototype phase (about nine months) for the spectrograph development was very enriching. It gave time to the different providers to refine their design, production plans and procedures, with what emerged during test phase. For AMU, responsible for the performance tests, the Fiber Illuminator Tool was already functional, but at the beginning, it required a lot of work to integrate it in the ICS. Then the analysis tools were set and optimized all along the process.

As presented in [14], the tests results lead to this first unit retrofitting, essentially mechanically, but also for some system parameters as the cryostats Linear Pulse Tubes (LPT) frequency or the shutters closing speed.

Critical parts delivery and qualification in the same area

The integration of all subsystems took place at Winlight Company in France. Winlight was responsible for the manufacturing and integration of most parts of the spectrograph and for the alignment of the optical spectrograph, the CEA in Saclay in France was responsible for the cryostats delivery including the CCD positioning inside them. The fact that these critical subsystems were done, integrated and tested in the same area (France) gave us much reactivity, with short time delays and no customs issues. We benefited from the swift reaction of the implicated teams for the delivery schedule, for instance with the inverted cryostat window lens issue (CEA and Winlight) and the collimator alignment issue (Winlight) (see section 3.6 for issues description).

Scientific tools development

The early calibration images were helpful for the development and optimization of the DRS. The software tools could be tested on all spectrographs with their diversity of optical performance and CCD cosmetics and behavior, and gain then in robustness and acuteness.

Furthermore, as already mentioned, some of them were extensively used for the qualification, especially the wavelength solution. Some more sophisticated tools (PSF modeling) helped the cryostats beat frequency detection issue (described in [14]).

All this scientist's work obviously benefits to the first operations at Mayall Telescope.

Time sparing for final installation at Mayall Telescope

The qualification work on each fully equipped spectrograph only cost about a month and a half on the whole schedule for a lot of benefits (taking into account cryostats integration, debugging plus performance test).

First it avoids late discover of issues despite compliant subsystems (gratings integration, window lens inverted, beat frequency between the LPT in the three cryostats... see section 3.6). Second it shortens integration and test at Telescope:

the spectrographs are already validated, the integration phase is shortened by the shorter focus ranges, and the DRS tools are mostly ready, the integration in ICS is facilitated.

Obviously some problems have to be solved, but the expertise acquired by all the teams helps for efficiency. For instance, a shipping problem was rapidly detected and solved during focus operation, thanks to the known compliance of the spectrograph.

Same test procedures

We used for the 10 spectrographs the same or nearly the same tests scripts for performing or analyzing their data. The results were very consistent which gave confidence in the design hypotheses and tolerances. This check is useful to the providers and project system team, for instance to propose a new similar instrument with confidence.

## 5.2 Limitations

Evolution of ICS and CCD configuration

We began the spectrograph qualification phase using some early versions of ICS and CCD configuration. Naturally, on these topics, the biggest effort was planned for later, for the final operation at Mayall Telescope. We suffered all along our work from successive systems incompatibilities or inadequate temporary CCD configuration files. The consequence was usually minor, just finding the right persons at the right moment, this induced small delays.

The test sparse slit: the first prototype

The spectrograph qualification phase was done with the first prototype of sparse fiber slit, which had some defects. But fortunately none of them was unacceptable, and the test sparse slit was compliant for our purpose. We took into account that some of the fibers were out of focus for the focus test. We took also advantage on one broken fiber to facilitate the image orientation check on the early times.

Possible missing analyses

Short schedules induced to shorten the test plan after some spectrographs validation and to orient the analyses to the main validation topics. Therefore, some second order problems can have been missed.

One tricky issue was not seen despite data was available: a bad spectral absorption feature in the blue arms as described in section 3.6: a collimator's coating performance deviation. This is for the moment the only issue which emerged.

## 6. CONCLUSIONS

The DESI spectrographs design, construction, integration, tests and installation at the Mayall have been very successful. Spectroscopic data meeting design performances were obtained at first light, and the DESI commissioning data have demonstrated the excellent stability of the spectrographs optics which is key for the quality of the spectral reduction. Given its performances, the DESI spectrograph design has been considered for upcoming spectroscopic surveys such as SDSS-V and the MegaMapper project.

## ACKNOWLEDGMENTS

This research is supported by the Director, Office of Science, Office of High Energy Physics of the U.S. Department of Energy under Contract No. DE–AC02–05CH1123, and by the National Energy Research Scientific Computing Center, a DOE Office of Science User Facility under the same contract; additional support for DESI is provided by the U.S. National Science Foundation, Division of Astronomical Sciences under Contract No. AST-0950945 to the NSF's National Optical-Infrared Astronomy Research Laboratory; the Science and Technologies Facilities Council of the United Kingdom; the Gordon and Betty Moore Foundation; the Heising-Simons Foundation; the French Alternative Energies and Atomic Energy Commission (CEA); the National Council of Science and Technology of Mexico; the Ministry of Economy of Spain, and by the DESI Member Institutions. The authors are honored to be permitted to conduct astronomical research on Iolkam Du'ag (Kitt Peak), a mountain with particular significance to the Tohono O'odham Nation. This work has been carried out thanks to the support of the OCEVU Labex (ANR-11-LABX-0060) and the A*MIDEX project (ANR-11-IDEX-0001-02) funded by the "Investissements d'Avenir" French government program managed by the ANR.